\magnification=\magstep1
\baselineskip = 15pt
\def\hp{Hess and Philipp}
\font\title = cmbx10 scaled 1440

\newcount\ftnumber
\def\ft#1{\global\advance\ftnumber by 1
          {\baselineskip=13pt \footnote{$^{\the\ftnumber}$}{#1 }}}
\newcount\fnnumber
\def\fn{\global\advance\fnnumber by 1
         $^{(\the\fnnumber)}$}
\def\fr#1/#2{{\textstyle{#1\over#2}}}

{\title
\centerline {Bell's theorem in the presence of}
\medskip
\centerline{classical communication} 
}
\bigskip
\centerline{N. David Mermin}
\centerline{Laboratory of Atomic and Solid State Physics}
\centerline {Cornell University, Ithaca, NY 14853-2501}

{\narrower \narrower \baselineskip = 12 pt

\bigskip

\noindent I explain what kinds of correlation or
even direct classical communication between detectors invalidate
Bell's theorem, and what kinds do not.

}

\bigskip

The immediate occasion for this note is the rejection by Karl Hess and
Walter Philipp [1] of my simple demonstration [2] that their
refutation [3] of an elementary version of Bell's theorem [4] cannot
be correct.  Rather than belabor misconceptions\ft{Among other
problems, their rejection of [2] relies on a misunderstanding of what
I mean by ``extended instruction sets''.} in [3], I prefer to examine
a question of more general interest.  Let us turn \hp\ upside down and
explore the extent to which Bell's theorem {\it survives\/}, not only
if, following \hp, we take advantage of properties of the detectors
correlated by the time on local synchronized clocks, but even if we
allow further correlation of the detectors through direct
straightforward ongoing classical communication between them.  How
much censorship must be imposed on the content of that communication
for Bell's theorem to remain valid?

The answer, which will surprise few who have thought much about Bell's
theorem, is that the theorem remains valid even if the detectors are
allowed to communicate with each other continuously throughout a long
series of runs, provided only that each detector is forbidden during
the course of each run from giving the other any information whatever
about the setting it has randomly been given in that run.  Aside from
that single necessary\ft{If information on the settings of the detectors
can be communicated before the detectors signal a result then the
quantum mechanical data (or any other data) can be trivially simulated
classically.} constraint the detectors can conspire in each run in any
way they like.\ft{Clearly detector responses that are driven by
non-communicating but internal computer programs, correlated only by
synchronized clocks, satisfy this constraint, so the proof below that
Bell's theorem survives such coordination of the detectors provides an
alternative demonstration to that in [2] that \hp\ are mistaken.}

To make things clear and simple I examine the question for the
particular geometry used in [2] and [4], but the argument can easily
be generalized.  In each of a long series of runs one of three
settings --- labeled 1, 2, or 3 --- is randomly and independently
assigned to each of the two detectors.  It is useful to introduce the
term ``wing'' to refer jointly to a particle and the detector that it
eventually arrives at.\ft{``Wings'' in the sense of wings of a
mansion: east wing, west wing, etc.  I borrow the term from
philosophers, who like to talk about the ``wings of the experiment''.}
The choice of setting for each wing is unknown to the other wing, and
the only constraint on the communication between wings is that each is
forbidden to reveal the value of its setting to the other.\ft{ We can
also allow the particles to communicate with each other (or with
either detector) even after they have left the source, provided, as
with the detectors, we forbid each particle from revealing to the
other wing any information it has discovered about the setting of its
own detector.}  Once the settings have been randomly, independently,
and secretly established for both wings and the wings have had any
further communications they wish --- always under the prohibition
against revealing their settings --- a light flashes red (R) or green
(G) in each wing and the run ends. The accumulated data in many runs
have two important features: (i) the lights flash the same colors
whenever the settings are the same; (ii) when the data are examined
without regard to the settings they are found to be quite random ---
in particular the same colors flash as often as different
colors.{\ft{Such data are produced by two spin-$\fr1/2$ particles in
the singlet state, when the detectors are Stern-Gerlach magnets, the
three settings are associated with measuring the spin along a
particular set of three coplanar directions 120$^\circ$ apart, and R
and G signal spin-up and spin-down at one detector, while signalling
spin-down and spin-up at the other.}

Can we construct a classical explanation for this data that respects
the fact that neither wing has any information about the setting in
the other wing when the lights flash?  The first feature of the data
to account for is (i), that the lights invariably flash the same
colors when the settings are the same.  With classical communication
there is no problem arranging for this without violating the
prohibition on revealing the settings. In every run, after each wing
has taken into account whatever conditions it might deem relevant, the
two wings agree on what color they will both flash for {\it each\/} of
the three possible settings.  The communication leading to such
agreement passes the censor because it reveals no information whatever
about the actual setting in each wing. The wings must negotiate such
an agreement in every run, whether or not the actual settings are the
same, because they do not know whether or not the settings are are the
same but do know that there is a $33\fr1/3$\% chance that they are.

But this essentially unique classical explanation of feature (i) of
the data cannot accomodate feature (ii), because it requires each run
of the experiment to be one of eight types:\ft{In earlier forms of Bell's
theorem the possibilities for coordinating behavior between the two
wings are much more constrained than they are here.  But even when the
two wings cannot directly talk to each other, they can still
characterize each run as one of these eight types and act accordingly,
which is essential for any classical explanation of feature (i), by
exploiting common information acquired by the particles before they
leave their common source.  \hp\ additionally emphasize the
possibility of using information available only at the detectors.}
those in which the agreed-upon colors to be flashed in either wing for
settings 1, 2, or 3 are RRG, RGR, GRR, GGR, GRG, RGG, RRR, or GGG.
The first six of these types each result in the same colors flashing
5/9 of the time.\ft{An RRG run, for example produces the same colors
for settings 11, 22, 33, 12, and 21, while it produces different
colors for settings 13, 23, 31, and 32; and the nine possibilities are
equally likely.}  The last two types result in the same colors always
flashing.  Therefore, no matter what colors are agreed upon in each
run, when all the data from all the runs are examined without regard
to what the actual settings were, the same colors will be found to
flash at least 5/9 of the time (Bell's inequality).  This contradicts
feature (ii) of the data, that the same colors flash only half the
time.  So no such classical explanation is possible (Bell's theorem).

This proof of Bell's theorem allows arbitrary communication between
the two wings, provided no information is revealed about the
settings.  The proof clearly allows variations in time of the
conditions of the detectors or the particles or both (independently),
which can be correlated either by synchronized clocks (as in the model
of \hp) or (in the more general case considered here) by direct
classical communication between the wings.  The proof allows
complex information to be available in either wing,
carried by either the particles or the detectors or both.  The
correlations allowed between the wings clearly include and go well
beyond the correlations envisaged by \hp.  But Bell's theorem continues
to hold.

\bigskip

\noindent {\sl Acknowledgment.}  Supported by the National Science 
Foundation, Grant No.~PHY0098429.

\bigskip
\centerline{\bf References}
\medskip
\parskip = 3pt
\parindent = 0pt

[1] Karl Hess and Walter Philipp, quant-ph/0207110.

[2] N. David Mermin, quant-ph/0206118.

[3] Karl Hess and Walter Philipp, quant-ph/0103028, Section 3.2. 

[4] N. David Mermin, Physics Today {\bf 38}, April 1985, p. 38. 

\bye